\documentclass[12pt]{article}

\usepackage{hyperref}

\usepackage{amsmath} 
\usepackage{amssymb}
\usepackage{amsfonts}
\usepackage{amsthm}
\usepackage{graphicx}

\newtheorem{theorem}{Theorem}

\newtheorem{proposition}{Proposition}

\theoremstyle{definition}
\newtheorem*{demo}{Proof}

\begin{document}
	\begin{center}
		\Large
	\textbf{Exact non-Markovian evolution with multiple reservoirs} \footnote{This work is supported by the Russian Science Foundation under grant~17-71-20154.}	
	
		\large 
		\textbf{A.E. Teretenkov}\footnote{Department of Mathematical Methods for Quantum Technologies,\\Steklov Mathematical Institute of Russian Academy of Sciences,
			ul. Gubkina 8, Moscow 119991, Russia\\ E-mail:\href{mailto:taemsu@mail.ru}{taemsu@mail.ru}}
		\end{center}
		
			\footnotesize
			The model of a multi-level system interacting with several reservoirs is considered. The exact reduced density matrix evolution could be obtained for this model without Markov approximation.
			Namely, this evolution is fully defined by the finite set of linear differential equations. In this work the results which were obtained previously for only one Lorentz peak in the spectral density are generalized to the case of an arbitrary number of such peaks. The case of the Ohmic contribution in the spectral density is also taken here into account.
			\normalsize

	\section{Introduction}
	
	The mathematically strict derivation of the Markovian equations for the reduced density matrix evolution originates in the work \cite{Bog39} by N.M. Krylov and N.N. Bogolyubov. The methods discussed there were developed in the theory of the stochastic limit. Its modern statement could be found in \cite{Accardi2002}. The master equation obtained by these methods have the \textit{Gorini--Kossakowski--Sudarshan--Lindblad} (GKSL) form \cite{Gorini76,Lindblad76}. 
	
	But problems of non-Markovian evolution have recently attracted  more and more interest \cite{Breuer99, Breuer07, Kossakowski07, ChruscinskiKossakowski10, Singh12, Tang13, Luchnikov19, Strathearn19}.  In particular, it is natural to ask the question, when the non-Markovian evolution of a system could be dilated to Markovian evolution of a finite-dimensional density matrix. In this work the model of a multi-level system interacting with reservoirs at zero temperatures is considered. The evolution of the density matrix for this multi-level system could be obtained exactly in terms of the finite-dimensional Shroedinger equation with the non-Hermitian Hamiltonian. We analyze the cases, when evolution of the reduced density matrix could be dilated to Markovian evolution of higher, but finite, dimension.
	
	This work develops and generalizes the results obtained in \cite{Teret19a} and \cite{Teret19b}.  In Sec.~\ref{sec:model} we describe our model and introduce the results of the work \cite{Teret19b}, which are necessary for this paper. The detailed discussion of connection of our model with other known models could be found in \cite{Teret19a} and \cite{Teret19b}. We just mention that this model is closely connected with the Friedrichs model\cite{Friedrichs48} and the approach, which we use in the propositions \ref{prop:LorPeaks} and \ref{prop:OhmicAndLorPeaks}, is a development of the pseudomode approach proposed in \cite{Garraway96, Garraway97, Garraway97a}. In Sec.~\ref{sec:LorPeaks} we consider the case, when the spectral density is the combination of the Lorentz peaks, and in Sec.~\ref{sec:OhmDen} we take into account the Ohmic contribution to the spectral density. Finally, in Conclusions we summarize our results and outline possible directions for the further studies.

	\section{Model}
	\label{sec:model}
	
 	We consider evolution in the Hilbert space
	\begin{equation*}
	\mathcal{H} \equiv (\mathbb{C}\oplus\mathbb{C}^{N}) \otimes \bigotimes\limits_{i=1}^N \mathfrak{F}_b(\mathcal{L}^2(\mathbb{R})).
	\end{equation*}
	Here $  \mathbb{C}\oplus\mathbb{C}^{N}$ is the $ (N+1) $-dimensional Hilbert space with a distinguished one-dimensional subspace. The degrees of freedom $ \mathbb{C}^{N} $ describe the excited states of the system and the distinguished subspace corresponds to the ground state.  Let $ | i \rangle , i = 0,  1, \ldots, N $ be the orthonormal basis of the space $  \mathbb{C}\oplus\mathbb{C}^{N}$, where $ | 0 \rangle $ corresponds to the distinguished subspace.  $ \mathfrak{F}_b(\mathcal{L}^2(\mathbb{R})) $ are bosonic Fock spaces corresponding to the reservoirs. Denote the vacuum state of these reservoirs by $ | \Omega \rangle $. Let us also introduce the creation and annihilation operators satisfying the canonical commutation relations: $ [b_{i}(k), b_{j}^{\dagger}(k')] = \delta_{ij} \delta (k - k')$, $[b_{i}(k), b_{j}(k')] = 0 $, $ b_{i}(k)| \Omega \rangle = 0$.
	
	We consider the system Hamiltonian of the general form with only the requirement that it vanishes on the ground state. Namely,
	\begin{equation}\label{eq:H_S}
	\hat{H}_S  = \sum_{i} \varepsilon_i |i \rangle \langle i| + \sum_{i\neq j} J_{ij} |i \rangle \langle j| =  0 \oplus H_S, \qquad i,j =1 ,\ldots,N,
	\end{equation}
	where we do not assume that $ \hat{H}_S $ is diagonalized in the basis $ |i \rangle  $. From the physical point of view $ |i \rangle  $ plays the role of the local basis \cite{Trushechkin16}. 
	
	The reservoir Hamiltonian is a sum of similar Hamiltonians of the free bosonic fields (with the same dispersion relation $ \omega(k) $)
	\begin{equation}\label{eq:H_B}
	\hat{H}_B =\sum_{i=1}^N \int \omega(k) b_{i}^{\dagger}(k)  b_{i}(k)  d k .
	\end{equation}
	
	The interaction is described by the following Hamiltonian
	\begin{equation}\label{eq:H_I}
	\hat{H}_I = \sum_i \int \left(  g^*(k)  | 0 \rangle \langle i| \otimes b_{i}^{\dagger}(k)+   g(k)  | i \rangle \langle 0 | \otimes b_{i}(k)\right) d k.
	\end{equation}
	Note that this expression takes into account that each level interacts with its own reservoir and the function $ g(k) $ (sometimes called form-factor \cite{Kozyrev17}) is the same for all the reservoirs. From the physical point of view such a Hamiltonian assumes that the dipole approximation (as only the terms, which are linear in creation and annihilation operators, are taken into account) and the rotating wave approximation (there are terms like $ | i \rangle \langle 0| \otimes b_{i}^{\dagger}(k) $ in the Hamiltonian) are satisfied.
	
	We consider the Sroedinger equation
	\begin{equation}\label{eq:Shr}
	\frac{d}{dt} | \Psi (t) \rangle = - i \hat{H} | \Psi (t) \rangle,
	\end{equation}
	with the Hamiltonian $ \hat{H} = \hat{H}_S \otimes I + I \otimes \hat{H}_B + \hat{H}_I $ and the initial condition
	\begin{equation}\label{eq:initCond}
	| \Psi (0) \rangle =( | \psi(0)\rangle + \psi_0(0) | 0 \rangle) \otimes | \Omega \rangle, \qquad \langle 0| \psi (0)\rangle = 0,
	\end{equation}
	i.e. the initial condition is fully factorized and assumes the zero temperature of the reservoir.
	
	We are interested in the evolution of the reduced density matrix
	\begin{equation*}
	\rho_S(t) \equiv \mathrm{tr}_R \; | \Psi (t) \rangle \langle \Psi (t)|,
	\end{equation*}
	where $ \mathrm{tr}_R \; $ is a partial trace with respect to reservoir, i.e. with respect to the Fock spaces $ \bigotimes\limits_{i=1}^N \mathfrak{F}_b(\mathcal{L}^2(\mathbb{R})) $. In~\cite{Teret19b} we have proved the following theorem which allows one to describe the evolution of the reduced density matrix in terms of the solution of the finite-dimensional integro-differential equation.
	\begin{theorem} \label{th:main}
		Let the integral (reservoir correlation function)
		\begin{equation}\label{eq:Gt}
		G(t) = \int |g(k)|^2 e^{-i \omega(k) t} dk
		\end{equation}
		converge for an arbitrary moment of time $ t \in \mathbb{R}_+ $ and define the continuous function, then
		\begin{equation*}
		\rho_S(t) =
		\begin{pmatrix}
		1 - || \psi (t) ||^2 & \psi_0(0) \langle \psi (t)|\\
		 \psi_0^*(0) | \psi (t) \rangle  & | \psi (t) \rangle   \langle \psi (t)|
		\end{pmatrix}
		\end{equation*}
		where $ | \psi (t) \rangle $ is the solution of the integro-differential equation
		\begin{equation}\label{eq:integroDiff}
		\frac{d}{dt} | \psi (t) \rangle = - i H_S| \psi (t) \rangle - \int_{0}^{t} ds \; G(t-s) | \psi (s) \rangle
		\end{equation}
		with the initial condition $ | \psi (t) \rangle|_{t=0} = | \psi (0) \rangle$.
	\end{theorem}
	
	In the physical literature the spectral density $ \mathcal{J}(\omega) $ is usually assumed to be given instead of the functions $ g(k) $ or $ G(t) $. The spectral density is the Fourier transform of $ G(t) $:
	\begin{equation*}
	G(t) = \int_{- \infty}^{+\infty} \frac{d \omega}{2 \pi} e^{- i \omega t}  \mathcal{J}(\omega), \qquad  \mathcal{J}(\omega) = \int_{- \infty}^{+\infty} G(t)  e^{i \omega t} dt.
	\end{equation*}
	In~\cite{Teret19b} we considered the case of the spectral density with only one Lorentz peak. In the next section we generalize these results for the case of an arbitrary finite number of Lorentz peaks and in Sec.~\ref{sec:OhmDen} we take into account the contribution of Ohmic spectral density.
	
	\section{Combination of Lorentz peaks}
	\label{sec:LorPeaks}
	
	In this section we reduce the evolution with spectral density which is a positive combination of Lorentz peaks
	\begin{equation}\label{eq:ourJtLor}
	\mathcal{J}_{L}(\omega) = \sum_{j=1}^K \frac{\gamma_j g_j^2}{ \left(\frac{\gamma_j}{2}\right)^2 + (\omega -\varepsilon_j)^2}, \qquad g_j>0, \gamma_i >0, K \in \mathbb{N},
	\end{equation}
	to a system of linear equations. The finite set of numbers $ g_j $ and the function $ g(k) $ are denoted by the same letter, because they have similar physical meaning. At the same time one should remember that $ g(k) $ is a complex-valued function and $ g_j $ are strictly positive real numbers. The correlation function for the spectral density \eqref{eq:ourJtLor} takes the form
	\begin{equation}\label{eq:ourGtLor}
	G_{\rm Lorentz}(t) = \sum_{j=1}^K g_j^2 e^{- \frac{\gamma_j}{2} |t| - i \varepsilon_j t}.
	\end{equation}
	We obtain the following proposition in this case.
	
	\begin{proposition}
		\label{prop:LorPeaks}
		Let $ | \psi (t) \rangle $ be a solution of Eq.~\eqref{eq:integroDiff} with the initial condition $ | \psi (t) \rangle|_{t=0} = | \psi (0) \rangle$ in the case, when $ G(t) = G_{\rm Lorentz}(t) $ is defined by formula \eqref{eq:ourGtLor}, then the $ (K+1)N $-dimensional vector $ | \tilde{\psi} (t) \rangle \equiv | \psi (t) \rangle \oplus  \oplus_{j=1}^K| \varphi_j (t) \rangle \in \oplus^{K+1}\mathbb{C}^N $, where 
		\begin{equation}\label{eq:pseudoDef}
		| \varphi_j (t) \rangle \equiv -i g_j \int_{0}^t ds \; e^{- \left(\frac{\gamma_j}{2} + i \varepsilon_j \right) (t-s)} | \psi (s) \rangle,
		\end{equation}
		satisfies the Schroedinger equation with the non-Hermitian Hamiltonian
		\begin{equation}\label{eq:nonHermEq}
		\frac{d}{dt} | \tilde{\psi} (t) \rangle = - i H_{\rm eff} | \tilde{\psi} (t) \rangle,
		\end{equation}
		\begin{equation*}
		H_{\rm eff} = 
		\begin{pmatrix}
		H_S & g_1 I_N & g_2 I_N &  \cdots & g_K I_N\\
		g_1 I_N & \left(\varepsilon_1 - i\frac{\gamma_1}{2} \right) I_N & 0 & \cdots & 0\\
		g_2 I_N & 0 &  \left(\varepsilon_2 - i\frac{\gamma_2}{2} \right) I_N & \cdots & 0\\
		\vdots & \vdots & \vdots & \ddots & 0\\
		g_K I_N &  0 & 0 & 0 & \left(\varepsilon_K - i\frac{\gamma_K}{2} \right) I_N\\
		\end{pmatrix},
		\end{equation*}
		with the initial condition $ | \tilde{\psi} (0) \rangle  = | \psi (0) \rangle  \oplus 0 $.
	\end{proposition}

	\begin{demo}
		Substituting \eqref{eq:ourGtLor} into \eqref{eq:integroDiff} and taking into account \eqref{eq:pseudoDef} we obtain
		\begin{equation*}
		\frac{d}{dt}| \psi (t) \rangle = - i H_S | \psi (t) \rangle - i \sum_{j=1}^K g_K | \varphi_j (t) \rangle.
		\end{equation*} 
		By differentiating \eqref{eq:pseudoDef} with respect to time $ t $ we also obtain $ K $ differential equations
		\begin{equation*}
		\frac{d}{dt} | \varphi_j (t) \rangle = - i g | \psi_j (t) \rangle - \left(\frac{\gamma_j}{2} + i \varepsilon_j \right) | \varphi_j (t) \rangle.
		\end{equation*}
		Combining these differential equations in the one differential equation for the vector $ | \tilde{\psi} (t) \rangle = | \psi (t) \rangle \oplus | \varphi_1 (t) \rangle \oplus \ldots \oplus | \varphi_K (t) \rangle = | \psi (t) \rangle \oplus  \oplus_{j=1}^K| \varphi_j (t) \rangle  $, we obtain \eqref{eq:nonHermEq}.  At the same time the definition \eqref{eq:pseudoDef} leads to $| \varphi_j (0) \rangle = 0 $, i.e. to the initial condition $ | \tilde{\psi} (0) \rangle  = | \psi (0) \rangle  \oplus 0 $. \qed
	\end{demo}

	In \cite{Teret19a} it was shown that if the matrix $ V = \frac{i}{2}(H_{\rm eff} - H_{\rm eff}^{\dagger}) $ is non-negative definite, then such a non-Hermitian Hamiltonian defines the (one-particle) GKSL equation such that $ \rho_S(t) $ could be obtained by the partial trace with respect to $ K N $ auxiliary degrees of freedom which have arisen in \eqref{eq:nonHermEq} in comparison with \eqref{eq:integroDiff}. Following \cite{Garraway97} we call such degrees of freedom pseudomodes and following \cite{Chruscinski17} we call the matrix $ V $ the optical potential. Without going into detail on this issue, let us note that
	\begin{equation*}
	V =  0 \oplus \frac{\gamma_1 }{2} I_N \oplus \cdots \oplus \frac{\gamma_K}{2}  I_N
	\end{equation*}
	and the fact that $ V $ is non-negative definite follows from the positivity of $ \gamma_j $. Thus, in the case, when the spectral density is a positive combination of Lorentz peaks, then in accordance with \cite{Teret19a} we have obtained that the evolution of the reduced density matrix  could always be dilated to Markovian (GKSL) evolution of higher dimension.
	
	\section{Ohmic spectral density}
	\label{sec:OhmDen}
	
	In this section we consider the influence of the Ohmic spectral density contribution
	\begin{equation*}
	\mathcal{J}_{\rm Ohmic}(\omega) = \eta \omega, \quad \eta > 0.
	\end{equation*}
	Since the Fourier transform of such a function exists only in the sense of generalized functions, then theorem \ref{th:main} is not directly applicable. Therefore, we consider the family of spectral densities with exponential cutoff
	 \begin{equation*}
	 \mathcal{J}_{\Omega}(\omega) = \eta \omega e^{- \frac{|\omega|}{\Omega}}
	 \end{equation*}
	parameterized by cutoff frequency $ \Omega $ which we tend to $ + \infty $. The corresponding reservoir correlation function has the following form
	\begin{equation}\label{eq:GOmega}
	G_{\Omega}(t) = -i \eta \frac{2 t \Omega^3}{\pi (1 +  (\Omega t)^2)}.
	\end{equation}
	We also need to consider the family of Hamiltonians
	\begin{equation}\label{eq:HSOmega}
	H_S(\Omega)=H_S^{(r)} + \frac{\eta \Omega}{\pi}.
	\end{equation}
	instead of $ H_S $. Further we will see that $ \frac{\eta \Omega}{\pi} $ plays the role of a counterterm \cite[Subsec.~3.6.1]{Breuer10}. As we tend $ \Omega \rightarrow + \infty $, then from the physical point of view form \eqref{eq:HSOmega} corresponds to the fact that the transition energies from the ground state to the excited state are much higher than the transition energies between the excited states.  Our research was motivated by transfer models in biological systems, where this condition is met \cite{Engel07, Lee07, Mohseni08, Plenio08}.  Moreover, it is this condition that validates neglecting multi-exciton states in our model described in Sec.~\ref{sec:model}.  
	
	\begin{proposition}
		Let $ | \psi_{\Omega} (t) \rangle $ be the solution of \eqref{eq:integroDiff} with the initial condition $ | \psi_{\Omega} (0) \rangle =| \psi (0) \rangle$, where $ G(t) = G_{\Omega}(t) + G_{\rm c}(t) $, $ G_{\Omega}(t) $ is defined by formula \eqref{eq:GOmega}, $ G_{\rm c}(t) $ is a continuous function, $ H_S = H_S(\Omega) $ is defined by formula \eqref{eq:HSOmega}. Let  the limit $ \lim_{\Omega \rightarrow + \infty} | \psi_{\Omega} (t) \rangle= | \psi^{(r)} (t) \rangle $ exist for $ t> 0 $  and define the infinitely differentiable function of time $ t \in (0, + \infty) $, then $ | \psi^{(r)} (t) \rangle  $ is a solution of the equation
		\begin{equation}\label{eq:regIntDiff}
		\frac{d}{dt}  | \psi^{(r)} (t) \rangle =  - i \frac{1}{1 + i \frac{\eta}{2}} H_S^{(r)}| \psi^{(r)} (t) \rangle - \int_{0}^{t} ds \;  \frac{1}{1 + i \frac{\eta}{2}}  G_c(t-s) | \psi^{(r)} (s) \rangle
		\end{equation}
		with the initial condition $ | \psi^{(r)} (0) \rangle = \frac{1}{1 + i \frac{\eta}{2}} | \psi (0) \rangle $.
	\end{proposition}

	\begin{demo}
	We give here only a brief sketch of the proof. Note that
	\begin{equation*}
	G_{\Omega}(t) = i \eta \frac{d}{dt} f_{\Omega}(t), \qquad f_{\Omega}(t) = \frac{1}{\pi} \frac{\Omega}{1 + (\Omega t)^2},
	\end{equation*}
	where $ f_{\Omega}(t)  $ is a symmetric $ \delta $ sequence. Then
	\begin{align*}
    -\int_{0}^{t} ds \;G_{\Omega}(t-s) | \psi_{\Omega} (s) \rangle &= - i \eta \int_{0}^{t} ds \; f_{\Omega}'(t-s) | \psi_{\Omega} (s) \rangle = i \eta \int_{0}^{t} ds \; \frac{d}{ds} f_{\Omega}(t-s) | \psi_{\Omega} (s) \rangle =\\
    &= i \eta  \left( \frac{\Omega}{\pi} | \psi_{\Omega} (t) \rangle - f_{\Omega}(t) | \psi (0) \rangle \right) -  i \eta \int_{0}^{t} ds \; f_{\Omega}(t-s) \frac{d}{ds} | \psi_{\Omega} (s) \rangle.
	\end{align*}
	By substituting this expression into the integral form of Eq.~\eqref{eq:integroDiff} we have
	\begin{align*}
	 | \psi_{\Omega} (t) \rangle - | \psi (0) \rangle = &- i H_S^{(r)} \int_0^t d \tau| \psi_{\Omega} (\tau) \rangle - \int_0^t d \tau\int_{0}^{\tau} ds \;G_c(\tau-s) | \psi_{\Omega} (s) \rangle -\\
	  &- i \eta \int_0^t d \tau f_{\Omega}(\tau) | \psi (0) \rangle-  i \eta \int_0^t d \tau \int_{0}^{\tau} ds \; f_{\Omega}(\tau-s) \frac{d}{ds} | \psi_{\Omega} (s) \rangle.
	\end{align*}
	(The terms of the form $ \int_0^t d \tau \frac{i \eta \Omega}{\pi} | \psi_{\Omega} (\tau) \rangle $ are canceled.) For $ t>0  $, $ \Omega \rightarrow + \infty $, we have
	\begin{align*}
	| \psi^{(r)} (t) \rangle  -| \psi (0) \rangle  = &- i H_S^{(r)} \int_0^t d \tau| \psi^{(r)} (\tau) \rangle - \int_0^t d \tau\int_{0}^{\tau} ds \;G_c(\tau-s) | \psi^{(r)} (s) \rangle -\\
	&- i  \frac{\eta}{2} | \psi (0) \rangle-  i \frac{\eta}{2} (| \psi^{(r)} (t) \rangle  -| \psi (0) \rangle ).
	\end{align*}
	Differentiating with respect to time and canceling $ 1+ \frac{\eta}{2} $ we obtain \eqref{eq:regIntDiff}. In addition, assuming in this equation $ t = 0  $ we obtain the required initial condition. \qed
	\end{demo}

 	Note that if one assumes $  G_c(t) = 0 $, i.e. the spectral density contains only the Ohmic term, then Eq.~\eqref{eq:regIntDiff} takes the form of Schroedinger equation \eqref{eq:nonHermEq} with the non-Hermitian Hamiltonian $ H_{\rm eff} = \frac{1}{1 + i \frac{\eta}{2}} H_S^{(r)} $. Corresponding optical potential has the form $ V  = \frac{\frac{\eta}{2}}{1 + \left(\frac{\eta}{2}\right)^2} H_S^{(r)}$. Thus, the condition that $ H_S^{(r)} $ is non-negative definite
 	\begin{equation}\label{eq:condForHS}
 	H_S^{(r)} \geqslant 0
 	\end{equation}
 	is necessary for possibility of Markovian dilation. From the physical point of view it means that the transition to the excited levels of the system Hamiltonian $ \hat{H}_S $ from the ground state corresponds to higher energies than $ \frac{\eta \Omega}{\pi} $ which is defined by cutoff frequency. 
 	
 	Analogously to Prop.~\ref{prop:LorPeaks} one could obtain the following proposition.
 	
	\begin{proposition}
	\label{prop:OhmicAndLorPeaks}	
	Let $ | \psi^{(r)} (t) \rangle $ be a solution of Eq.~\eqref{eq:regIntDiff} with the initial condition $ | \psi (t) \rangle|_{t=0} = | \psi (0) \rangle$ in the case, when $ G_c(t) = G_{\rm Lorentz}(t)  $ is defined by formula \eqref{eq:ourGtLor}, then the $ (K+1)N $-dimensional vector $ | \tilde{\psi} (t) \rangle \equiv | \psi (t) \rangle \oplus  \oplus_{j=1}^K| \varphi_j (t) \rangle \in \oplus^{K+1}\mathbb{C}^N $, where $ | \varphi_j (t) \rangle  $ is defined similarly to \eqref{eq:pseudoDef}, satisfies the Schroedinger equation \eqref{eq:nonHermEq} with the non-Hermitian Hamiltonian
		\begin{equation*}
		H_{\rm eff} = 
		\begin{pmatrix}
		\frac{1}{1 + i \frac{\eta}{2}} H_S^{(r)} & \frac{g_1}{1 + i \frac{\eta}{2}}  I_N & \frac{g_2}{1 + i \frac{\eta}{2}}  I_N &  \cdots & \frac{g_K}{1 + i \frac{\eta}{2}}  I_N\\
		g_1 I_N & \left(\varepsilon_1 - i\frac{\gamma_1}{2} \right) I_N & 0 & \cdots & 0\\
		g_2 I_N & 0 &  \left(\varepsilon_2 - i\frac{\gamma_2}{2} \right) I_N & \cdots & 0\\
		\vdots & \vdots & \vdots & \ddots & 0\\
		g_K I_N &  0 & 0 & 0 & \left(\varepsilon_K - i\frac{\gamma_K}{2} \right) I_N\\
		\end{pmatrix}.
		\end{equation*}
	\end{proposition}
	
	In order to examine if optical potential is non-negative definite in this case, let us note that if one moves in each subspace $ \mathbb{C}^N $  of the space $ \oplus^{K+1}\mathbb{C}^N  $ into the global basis (the eigenbasis of $ H_S $), then $ H_{\rm eff} $ could be decomposed into the direct sum of the blocks in the following way ($ E_{\alpha} $~are eigenvalues of $ H_S  $)
	\begin{equation*}
	H_{\rm{eff}, \alpha}  = \begin{pmatrix}
	\frac{1}{1 + i \frac{\eta}{2}} E_{\alpha} & \frac{1}{1 + i \frac{\eta}{2}}  \textbf{g}^T\\
	\textbf{g} & \varepsilon - i \frac12\gamma
	\end{pmatrix}, \quad \textbf{g}^T = (g_1 \ldots g_K), \varepsilon = \mathrm{diag} \; (\varepsilon_1, \cdots, \varepsilon_K), \gamma = \mathrm{diag} \; (\gamma_1, \cdots, \gamma_K).
	\end{equation*}
	Then the optical potential could be decomposed into the direct sum of the blocks in the following way
	\begin{equation*}
	V_{\alpha} =\frac{i}{2}(H_{\rm{eff}, \alpha} - H_{\rm{eff}, \alpha}^{\dagger}) = 
	\begin{pmatrix}
	\frac{\frac{\eta}{2}}{1 + \left(\frac{\eta}{2}\right)^2} E_{\alpha} & \frac{\frac{\eta}{2}}{2(1 + i \frac{\eta}{2})}  \textbf{g}\\
	\frac{\frac{\eta}{2}}{2(1 + i \frac{\eta}{2})} \textbf{g}^T & \frac12\gamma
	\end{pmatrix}.
	\end{equation*}
	Since the submatrix $ \frac12\gamma $ is positive definite, then it is sufficient to test if the determinant  
	\begin{equation*}
	\det V_{\alpha}
	=
	\left(\frac{\frac{\eta}{2}}{1 + \left(\frac{\eta}{2}\right)^2} E_{\alpha} - \frac{\left(\frac{\eta}{2}\right)^2}{2(1 + \left(\frac{\eta}{2}\right)^2)} \textbf{g}^T \gamma^{-1} \textbf{g} \right) \det \left(\frac12\gamma\right).
	\end{equation*}
	is positive to show that $ V_{\alpha} $ is positive definite. The condition $ \det V_{\alpha}>0 $ is equivalent to
	\begin{equation*}
	E_{\alpha} > \frac{\eta}{4} \textbf{g}^T \gamma^{-1} \textbf{g} = \frac{\eta}{4} \sum_{j=1}^K \frac{g_j^2}{\gamma_j}.
	\end{equation*}
	In general one should examine non-negativity of all the diagonal minors of a matrix to show that this matrix is non-negative definite \cite[Ch. 10, \S 4]{Gantmacher04}. However, all the minors containing the element $ \frac{\frac{\eta}{2}}{1 + \left(\frac{\eta}{2}\right)^2} E_{\alpha}  $ have the form similar to the determinant of $ V_{\alpha} $. But they contain all possible vectors consisting of subsets of $ \textbf{g} $ instead of full vector $ \textbf{g} $. Hence, if 
	\begin{equation*}
	E_{\alpha} \geqslant \frac{\eta}{4} \sum_{j=1}^K \frac{g_j^2}{\gamma_j},
	\end{equation*}
	then the inequalities for all other minors are satisfied automatically due to the fact that $ \frac{g_j^2}{\gamma_j} $ is positive. This expression (the condition of Markovian dilation possibility) could be rewritten in the matrix form
	\begin{equation*}
	H_S^{(r)} \geqslant \left(\frac{\eta}{4} \sum_{j=1}^K \frac{g_j^2}{\gamma_j} \right) I_N,
	\end{equation*}
	which is the generalization of condition \eqref{eq:condForHS} for the case $  G_c(t) = 0 $. From the physical point of view it corresponds to the fact that the transition energies to the excited levels should be greater than the sum of the counterterm which is proportional to the cutoff frequency and the term which is proportional to the sum of the Lorentz peaks maxima.
	
\section{Conclusions}

In this work the exact evolution of the reduced density matrix is obtained for the model described in Sec.~\ref{sec:model}. It is expressed in terms of the solution of the Schroedinger equation with the non-Hermitian Hamiltonian. The cases, when the spectral density is a combination of Lorentz peaks as well as when there is the Ohmic contribution to the spectral density, is considered. The conditions, when evolution of the reduced density matrix could be dilated to the Markovian one, are obtained.

As a possible direction for further study, we find it interesting to try to obtain the exact non-Markovian evolution described in this article as the limit of the approximate evolution of a more general kind. Just as the Markovian equations could be obtained the Bogolyubov -- Van Hove limit \cite[Sec~1.8]{Accardi2002}. 

\section{Acknowledgments}

The author expresses his deep gratitude to I.\,V.~Volovich, S.\,V.~Kozyrev, A.\,I.~Mikhailov and A.\,S.~Trushechkin for the fruitful discussion of the problems considered in the work.


\begin{thebibliography}{99}
	
	\bibitem{Bog39}
	\emph{Krylov N.M., Bogolubov N.N.}
	On derivation of Fokker-Planck equations by perturbation-theory method based on spectral properties of the perturbation Hamiltonian, // Zap. Kaf. Mat. Fiz. Ukrain. Akad. Nauk 1939. V. 4. P. 5--80 [in Ukrainian].
	
	\bibitem{Accardi2002}
	\emph{Accardi L., Lu Y.G., Volovich I.} Quantum theory and its stochastic limit. Berlin:Springer, 2002.
	
	\bibitem{Gorini76}
	\emph{Gorini V., Kossakowski A., Sudarshan E. C. G.} Completely positive dynamical semigroups of N-level systems // J. of Math. Phys. 1976. V. 17, N 5. P. 821-825.
	
	\bibitem{Lindblad76}
	\emph{Lindblad G.} On the generators of quantum dynamical semigroups // Comm. in Math. Phys. 1976. V. 48, N 2. P. 119-130.
	
	\bibitem{Breuer99}	
	\emph{Breuer~H.\,P., Kappler~B., Petruccione~F.} Stochastic wave-function method for non-Markovian quantum master equations // Phys. Rev. A. 1999. V. 59, N 2. P. 1633.
	
	\bibitem{Breuer07}
	\emph{Breuer~H.\,P.} Non-Markovian generalization of the Lindblad theory of open quantum systems // Phys. Rev. A. 2007. V. 75, N 2. P. 022103.
	
	\bibitem{Kossakowski07}
	\emph{Kossakowski~A., Rebolledo~R.} On non-Markovian time evolution in open quantum systems // Open Systems and Information Dynamics. 2007. V. 14, N 3. P. 265--274.
	
	\bibitem{ChruscinskiKossakowski10}
	\emph{Chruscinski~D., Kossakowski~A.} Non-Markovian quantum dynamics: local versus nonlocal // Phys. Rev. Lett. 2010. V. 104, N 7. P. 070406.
	
	\bibitem{Singh12}
	\emph{Singh~N., Brumer~P.} Efficient computational approach to the non-Markovian second order quantum master equation: electronic energy transfer in model photosynthetic systems // Mol. Phys. 2012. V. 110, N 15--16. P. 1815--1828 .
	
	\bibitem{Tang13}
	\emph{Tang~N., Xu~T.-T., Zeng~H.-S.} Comparison between non-Markovian dynamics with and without rotating wave approximation // Chinese Phys. B. 2013. V. 22, N 3, P. 030304.
	
	\bibitem{Luchnikov19}
	\emph{Luchnikov~I.A., Vintskevich~S.V., Ouerdane~H., Filippov~S.N.} Simulation Complexity of Open Quantum Dynamics: Connection with Tensor Networks // Phys. Rev. Lett. 2019. V. 122, N 16.
	
	\bibitem{Strathearn19}
	\emph{Strathearn A. et al.} Efficient non-Markovian quantum dynamics using time-evolving matrix product operators // Nature Comm. 2018. V. 9, N 1.  P. 3322.
	
	
	\bibitem{Teret19a}
	\emph{Teretenkov A.E.} Pseudomode Approach and Vibronic Non-Markovian Phenomena in Light-Harvesting Complexes // Proc. Steklov Inst. Math. 2019. V. 306. P. 242--256.
	
	\bibitem{Teret19b}
	\emph{Teretenkov A.E.} Non-Markovian Evolution of Multi-level System Interacting with Several Reservoirs. Exact and Approximate // Lobachevskii J. of Math. 2019. V. 40, N. 10. P. 1587-1605
	
	\bibitem{Friedrichs48}
	\emph{Friedrichs K.O.} On the perturbation of continuous spectra // Comm. on Pure and Applied Math. 1948. V. 1, N 4. P. 361-406
	
	\bibitem{Garraway96}
	\emph{Garraway B.M., Knight P.L.} Cavity modified quantum beats // Phys. Rev. A. 1996. V. 54, N~4. P. 3592.
		
	\bibitem{Garraway97}	
	\emph{Garraway B.M.} Nonperturbative decay of an atomic system in a cavity // Phys. Rev. A. 1997. V. 55, N 3. P. 2290.
		
	\bibitem{Garraway97a}		
	\emph{Garraway B.M.} Decay of an atom coupled strongly to a reservoir // Phys. Rev. A. 1997. V. 55, N 6. P. 4636.
		
	\bibitem{Trushechkin16}
	\emph{Trushechkin A.S., Volovich I.V.} Perturbative treatment of inter-site couplings in the local description of open quantum networks // EPL. 2016. V. 113, N 3. P. 30005.
	
	\bibitem{Chruscinski17}
	\emph{Chruscinski~D., Pascazio~S.} A brief history of the GKLS equation // arXiv preprint arXiv:1710.05993. 2017.
	
	\bibitem{Kozyrev17}
	\emph{Kozyrev S.V. et al.} Flows in non-equilibrium quantum systems and quantum photosynthesis // Inf. Dim. Anal., Quant. Prob. and Rel. Top. 2017. V. 20, N 4. P. 1750021.
	
	\bibitem{Breuer10}
	Breuer H.-P., Petruccione F. \emph{The theory of open quantum systems}, Oxford: Oxford University Press, 2002.
	
	\bibitem{Engel07}
	\emph{Engel~G.\,S.~et al.} Evidence for wavelike energy transfer through quantum coherence in photosynthetic systems // Nature. 2007. V. 446, N 7137. P. 782.
	
	\bibitem{Lee07}
	\emph{Lee~H., Cheng~Y.C., Fleming~G.R.} Coherence dynamics in photosynthesis: protein protection of excitonic coherence // Science. 2007. V. 316, N 5830. P. 1462--1465.
	
	\bibitem{Mohseni08}
	\emph{Mohseni~M. et al.} Environment-assisted quantum walks in photosynthetic energy transfer //  The J. of Chem. Phys. 2008. V. 129, N 17. P. 11B603.
	
	\bibitem{Plenio08}
	\emph{Plenio~M.B., Huelga~S.F.} Dephasing-assisted transport: quantum networks and biomolecules //  New J. of Phys. 2008. V. 10, N 11, P. 113019.
	
	\bibitem{Gantmacher04}
	Gantmacher~F.\,R. \emph{Matrix theory}. 5th ed. Moscow: FISMATLIT, 2004 [in Russian].
	
\end{thebibliography}
\end{document}